\documentclass[12pt,a4paper]{article}
\usepackage[textwidth=450pt, textheight=660pt]{geometry}
\usepackage{amsmath,amsthm,amssymb}
\usepackage{graphicx,tabu}
\usepackage[pagebackref=false]{hyperref}
\usepackage[active]{srcltx}
\usepackage[affil-it]{authblk}
\usepackage{wrapfig}
\usepackage[noadjust]{cite}
\usepackage{filecontents}
 
 \newcommand{\be}{\begin{equation}}
\newcommand{\ee}{\end{equation}}
\newcommand{\bea}{\begin{eqnarray}}
\newcommand{\eea}{\end{eqnarray}}
\newcommand{\nn}{\nonumber}
\newcommand{\tr}{\textrm{Tr}}
\newcommand{\trm}{\textrm}
\newcommand{\tit}{\textit}
\newcommand{\tsf}{\textsf}
\newcommand{\tbf}{\textbf}
\newcommand{\bs}{\boldsymbol}
\newcommand{\ba}{\begin{array}}
\newcommand{\ea}{\end{array}}
\newcommand{\bfig}{\begin{figure}}
\newcommand{\efig}{\end{figure}}

\begin{document}

  \vspace{2cm}

  \begin{center}
    \font\titlerm=cmr10 scaled\magstep4
    \font\titlei=cmmi10 scaled\magstep4
    \font\titleis=cmmi7 scaled\magstep4
  {\bf Fermionic Casimir  effect  in  Graphene }

    \vspace{1.5cm}
     \noindent{{\large Y. Koohsarian ${}^a$ \footnote{yo.koohsarian@mail.um.ac.ir},
  A. Shirzad  ${}^{b,c}$ \footnote{shirzad@ipm.ir}}} \\
     ${}^{a}$ {\it Department of Physics, Ferdowsi University of Mashhad \\
       P.O.Box  91775-1436, Mashhad, Iran} \\
   ${}^b$ {\it Department of Physics, Isfahan University of Technology \\
       P.O.Box 84156-83111, Isfahan, Iran,\\
      ${}^c$  School of Physics, Institute for Research in Fundamental Sciences (IPM),\\
       P. O. Box 19395-5531, Tehran, Iran} \\

  \end{center}
  \vskip 2em

\begin{abstract}
 We investigate the Fermionic Casimir effect at finite temperature  for two parallel chain of adatoms in a Graphene sheet, and the corresponding Casimir force is interpreted as an interaction between the adatom chains. We apply useful techniques to find asymptotically explicit expressions for the Casimir energy, for small as well as large temperatures (with respect to the effective temperature of the Graphene). We obtain a value in the order of $10^{-2} N/m$ for the Casimir force between  (per unit length of) the adatom chains being one nanometer apart, which is considerably  noticeable e.g. in comparison to the experimental values  for  built-in tension of a suspended Graphene sheet.
\end{abstract}

 \textbf{Keywords} \\  Fermionic Casimir effect, Graphene sheet,  Adatoms, Built-in tension.

\section{Introduction} \label{sec-1}
 Casimir effect, as an important macroscopic manifestations of the zero-point quantum oscillations \cite{Cas}, nowadays has been investigated extensively for various configurations \cite{CR1,CR2,CR3,CR4,CR5,CR6}, and is expected to play  a significant role in submicron dimensions \cite{CNR1,CNR2,CNR3}.  Graphene, a single atomic layer of Carbon atoms  covalently bounded in a hexagonal lattice, as the thinnest known material with remarkable and exceptional properties \cite{G1,G2}, is considered nowadays as an ideal candidates for constructing submicroscale systems, see \cite{GR1,GR2,NEMS1,NEMS2,NEMS3} as reviews for various aspect and applications of Graphene. Hence investigating the Casimir effect for the Graphene, can have considerable importance e.g. for these ultra-miniature systems. 
 
 One of the most interesting features of the Graphen is the  low-energy  behavior of its electronic excitations: the electron-hole excitations of Graphene at  low energy, behave as massless Dirac fermions in a $2$-dimensional space, with the light-velocity replaced by the Fermi-velocity \cite{GD}. In this paper we investigate the fermionic Casimir effect originating from the zero-point energy of these  Dirac-like fermions of the Graphene. Such a Casimir-like interactions has been generally predicted  in Ref. \cite{IA}, by calculating the interaction energy of the electron scattering between (single) adatoms in Graphene, using the known tight-binding Hamiltonian, in which, the presence of adatoms has been described by an additional potential term. Here we use the massless Dirac Hamiltonian to calculate the Casimir energy between adatoms.  In fact, as we know, the mentioned  electron-hole excitations are induced by  hopping $\pi$-electrons of Graphene, hence consuming these $\pi$-electrons   by e.g. some adatoms on Graphene, can break the isotropy of the $\pi$-electron distribution, giving rise to a net interaction between adatomes. So, in other words, taking these adsorbates as  some external restricting boundaries on the Dirac-like field in the Graphene, the resulting  Casimir forces  can be interpreted as  interactions between the adatoms. Note that this is different from the usual Casimir effect  as an interaction between  two Graphene sheets (or e.g. between a Graphene sheet  and  a substrate) which is arisen actually from the  zero-point  oscillations of the   electromagnetic field  between the Graphene sheets, see \cite{CEMG1, CEMG2, CEMG3,CEMG4}. Here, the fermionic Casimir energy of the Graphene comes from the zero-point oscillation energy of a  Dirac-like  spinor field having the Graphene electron-hole as its quantum particle states (similar to the Photons for the quantized electromagnetic field). We first find the   zero-point energy of  Dirac spinors at finite  temperature, by applying the conventional imaginary-time formalism. Then, as a result, we would be able to calculate the Casimir energy   between  two parallel chain of adatoms, in a (monolayer) Graphene sheet.  The boundary conditions on the  Dirac-like fields in the Graphene sheet would be given by the familiar \textit{bag} boundary conditions on the adatom-chain lines. We use the known Riemann and Hurwitz zeta functions to regularize the divergent zero-point sums, and apply some new useful techniques to obtain asymptotically explicit expressions for the Casimir force for  small as well as  large effective-temperatures, with respect to the room temperature. We compare the values of the Casimir forces with the experimental values, given in Refs. \cite{GT1,GT2,GT3}, for the pre-tensions of the Graphene at room temperature.
  
  \section{ Zero-point energy of  Dirac fermions at finite temperature } \label{sec-2}
  
  From the functional formalism of the quantum field theory, the  zero-point energy of a field can be written (up to a total-time factor which would be automatically canceled in the forthcoming calculations)  in the form \cite{IQFT,CR3}
  \be
E_0 \sim i \ln{Z_0} \label{eq-1}
\ee
 with $Z_0$ as the generating functional in the absence of any source, and we have used the Planck unites. For massless Dirac fermions we simply  have (up to an irrelevant constant)
\bea
Z_0 \sim  \det  \left( \gamma^{\mu} \partial_{\mu} \right), \label{eq-2}
\eea
in which  $\gamma^{\mu}$'s are the Dirac gamma matrices, and so the zero-point energy \eqref{eq-1} for massless Dirac fermions takes the form
  \bea
E_0 \sim i \tr \ln{\left( \gamma^{\mu} \partial_{\mu} \right)} \label{eq-3}
\eea
where the trace is taken on a Dirac $4$-spinor.  Here in order to simplify the forthcoming  calculations, we rewrite the above equation as
  \bea
E_0 &\sim& \frac{i}{\tsf{2}} \tr \ln{\left( \gamma^{\mu} \partial_{\mu} \right)^2} \nn \\
 &\sim& 2i \tr  \ln{ \left(\partial^2 \right)}
 \label{eq-4}
\eea  
  where,  the second line has been multiplied by $4$, since the trace in the second line would be taken on only one component of the Dirac $4$-spinor. Then, in the presence of appropriate external boundaries on the  Dirac field,  the mode solutions would be discretized,   and so  the zero-point energy \eqref{eq-4} takes the form 
 \be
E_0= 2i \int  \frac{d\omega}{2\pi}  \sum_J   \ln\left[-\omega^2+ k_J^2 \right]. \label{eq-5}
 \ee
where,  $\omega$ and $k$ represent the energy and the momentum of the Dirac modes, and  the collective index $J$ labels the mode numbers,  and here the total-time factor would have been automatically canceled. One can show that the above equation is equivalent to the familiar form,
   \be
E_0= -2  \sum_J    k_J,  \label{eq-6}
 \ee
  for   the zero-point energy  of  the Dirac spinors. To show  this, first we rewrite Eq. \eqref{eq-5} as  \cite{CR3}
   \bea
E_0 &=&  2 i \sum_J  \int_{-\infty}^{\infty}\frac{d\omega}{2\pi} \lim_{s \rightarrow 0} \left(-\frac{\partial }{ \partial s}\right)  \left(-\omega^2 + k_J^2 \right)^{-s}  \nn \\
 &=& -2 i  \lim_{s \rightarrow 0} \frac{\partial }{ \partial s}\sum_J k_J^{-s} \int_{-\infty}^{\infty}\frac{d z}{2\pi}  \left(1- z^2 \right)^{-s},  \label{eq-7} 
 \eea
where in the second line, we have introduced a new variable $\omega=z k_J$. Then using
\bea
\int_{-\infty}^{\infty}dz\left(1-z^2 \right)^{-s} = i \sqrt{\pi}  \frac{\Gamma \left(s-\frac{1}{2}\right)}{\Gamma(s)}, \nn
\eea
 see e.g. \cite{math4}, we find
 \bea
E_0=  \frac{1}{\sqrt{\pi}}  \lim_{s \rightarrow 0} \frac{\partial }{ \partial s} \frac{\Gamma \left(s-\frac{1}{2}\right)}{\Gamma(s)}\sum_J k_J^{-s} , \nn
 \eea
which equals  just  to the zero-point energy in  Eq.  \eqref{eq-6}. 

 Now,  to find the zero-point energy  at finite temperature, we use the known imaginary-time formalism, in which, the temperature is introduced  by a periodical condition on the field in the imaginary-time,  through which, the temperature would appear in the  Matsubara frequencies, see e.g. \cite{CR1,CR2,CR3,CR4}. Since,  through the  the above calculations, we have not used the  imaginary version for the time variable, here equivalently we can use an  imaginary version for  the Matsubara frequencies
\be
  -i\omega_l=2\pi l T \ \ \ ; \ \ \ l=0,\pm1,\pm2,... \label{eq-8}
\ee
 with $T$ as the temperature. As a a result, the zero-point energy  of the massless Dirac field \eqref{eq-5}, at finite temperature  takes the form 
 \be
E_0(T) = -2T \sum_J \sum_{l=-\infty}^{\infty}   \ln\left[(2\pi T)^2 l^2+ k_J^2 \right]. \label{eq-9}
 \ee
Obviously for massive fermions, a squared mass term should  be included in the  logarithm argument.    Note that, except for a coefficient ($-4$), the finite-temperature zero-point energy \eqref{eq-9} for the spinor field, is equivalent to the well known zero-point energy of  scalar field at finite temperature (see e.g. \cite{CR3}), where the minus sign of the coefficient is a characteristic feature of the fermionic field, and the factor $4$, turns up on account of the four component of the Dirac spinor, as previously mentioned.

\section{Casimir effect for quasi-spinors in Graphene } \label{sec-2}

It has been demonstrated that the low-energy behavior  of  electronic excitations of the Graphene can be described by a massless  Dirac-like equation in a 2-dimensional space, with the light velocity $c$ replaced by the Fermi velocity $ \upsilon_{\trm{F}} \approx c/300$, see \cite{GD}. As a result, the zero-point energy \eqref{eq-4}, for the quasi-spinors of Graphene can be directly written as
 \bea
E_0 = 2i \hbar  \ \tr  \ln{ \left(\frac{\partial_t^2}{\upsilon_{\trm{F}}^2}- \partial_x^2-\partial_y^2 \right)}
 \label{eq-10}
\eea 
 in which $(t, x, y)$  represent a (2+1) space-time, and we have restored the Planck constant $\hbar$. Now to find the discretized momentums, we must apply appropriate boundary condition on the  Graphene quasi-spinors. But, as we previously discussed, the adatoms on Graphene  can play the role of  some external restricting boundaries on the Graphene Dirac-like field. Here we consider the configuration of figure \ref{AP},  i.e.  two parallel chain of adatoms  in a Graphene sheet. These parallel  adatom-chains can be regarded as a 1-dimensional version of  the parallel planes considered in Refs. \cite{bag,CR3,CR6}. The mode solutions of  a 2-dimensional Dirac equation, for the configuration of these adatom-chains,  can be written as
  \bea
 \psi_{k_x,k_y}(x, y, t)= e^{-i \omega  t } e^{i  k_y y} \left( u e^{i k_x x} + v e^{-i  k_x x} \right),
 \label{eq-11}
 \eea
in which $u$ and $v$ are constant spinors.  But the  appropriate boundary condition for Dirac spinors is given by the known \tit{bag} boundary condition \cite{CR6,bag}, which  is equivalent to requiring the normal component of the fermion current to vanish at the boundaries \cite{CR3}, or in other word, supposing the fermion current could not penetrate through the boundaries, as physically expected. Now considering the configuration of figure \ref{AP}, the bag boundary condition for the mode solutions \eqref{eq-11}, can be written as
\bfig [!]
\centering
\includegraphics[width=10 cm, height=3 cm]{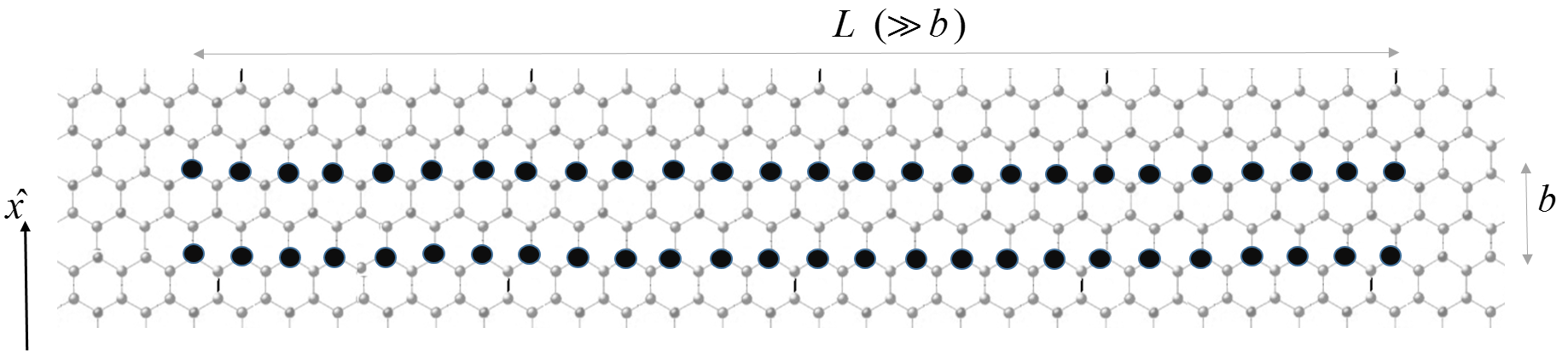}
\caption{Two parallel chain of adatoms (black pellets) on a Graphene sheet}
\label{AP}
 \efig
 
 \bea
  \left[i \bs{\gamma} \cdot  \hat{\tbf{n}} +1 \right] \psi(x,y) |_{x=0,a}=0 
    \label{eq-12}
 \eea
where $\hat{\tbf{n}}$ is  a unite vector normal to the boundaries, having $\hat{\tbf{n}}=\hat{\tbf{-x}},\hat{\tbf{x}}$ for  $x=0,a$ respectively, and $\bs{\gamma}$ is  the spacelike component of the Dirac gamma matrices.  Now  applying the boundary conditions \eqref{eq-12} onto the solution \eqref{eq-11}, and after some calculations similar to those of Ref. \cite{CR6}, one can find the discretized modes with
 \be
 k_{x,n}=\left(n+\frac12\right) \frac{\pi}{a} ; \ n=0,1,2,...
 \label{eq-13}
 \ee
As a result, the zero-point energy  \eqref{eq-5} takes the form
 \be
E_0= 2i \int_{-\infty}^{\infty}  \frac{d\omega}{2\pi} \int_{-\infty}^{\infty}  \frac{L}{2\pi} d k_y  \sum_{n=0}^{\infty}  \ln\left[-\frac{\omega^2}{\upsilon_{\trm{F}}^2}+ k_y^2+ \left(n+\frac12\right)^2 \frac{\pi^2}{a^2} \right]. 
\label{eq-14}
 \ee
In which, $L$ is the length of the adatom-chains ($L\gg a$), see Fig. \ref{AP}. Then using the Matsubara frequencies \eqref{eq-8}, the zero-point energy for the configuration of two parallel adatom-chains, can be written as 
 \be
E_0(a,T) = -2 k_B T \sum_{l=-\infty}^{\infty} \int_{-\infty}^{\infty} \frac{L}{2\pi} d k_y  \sum_{n=0}^{\infty}  \ln\left[\left(\frac{2\pi k_B T}{\hbar \upsilon_{\trm{F}}}\right)^2  l^2+ k_y^2+ \left(n+\frac12\right)^2 \frac{\pi^2}{a^2} \right]. 
\label{eq-15}
 \ee
with  $k_B$ as  the Boltzmann constant, and  $T$ as the Graphene temperature. In order to regularize the above zero-point energy, we first write it as a parametric integral;
 \bea
E_0(a,T)& =& 2 k_B T  \sum_{l=-\infty}^{\infty}  \int_{-\infty}^{\infty} \frac{L}{2\pi} d k_y   \sum_{n=0}^{\infty}\lim_{s\rightarrow 0} \frac{\partial}{\partial s}  \delta^{-2s} \left[\left(\frac{2\pi k_B T}{\hbar \upsilon_{\trm{F}}}\right)^2  l^2+ k_y^2+ \left(n+\frac12\right)^2 \frac{\pi^2}{a^2} \right]^{-s} \nn \\
&=&  2 k_B T \sum_{l=-\infty}^{\infty}  \int_{-\infty}^{\infty}  \frac{L}{2\pi} d k_y  \sum_{n=0}^{\infty}\lim_{s\rightarrow 0} \frac{\partial}{\partial s}  \delta^{-2s} \nn \\
 && \hspace{1 cm} \times \int_0^{\infty} \frac{dt}{t} \frac{t^s}{\Gamma(s)} \exp \left(-t\left[\left(\frac{2\pi k_B T}{\hbar \upsilon_{\trm{F}}}\right)^2  l^2+ k_y^2+ \left(n+\frac12\right)^2 \frac{\pi^2}{a^2} \right]\right) 
 \label{eq-16}
\eea
 where $\delta$  is an arbitrary parameter with the dimension of the length,  introduced for dimensional requirements. Taking a Gaussian integral for $k_y$, the above equation turns to
 \bea
&&E_0(a,T)= 2 L \sqrt{\pi}\frac{(k_B T)^2}{\hbar \upsilon_{\trm{F}}} \sum_{l=-\infty}^{\infty}   \sum_{n=0}^{\infty}\lim_{s\rightarrow 0} \frac{\partial}{\partial s}  \left(\frac{\hbar \upsilon_{\trm{F}}}{2\pi  k_B T\delta}\right)^{2s}  \nn \\
 && \hspace{4 cm} \times \int_0^{\infty} \frac{dt}{t} \frac{t^{s-1/2}}{\Gamma(s)} \exp \left(-t\left[ l^2 +\lambda_a^2 \left(n+\frac12\right)^2   \right]\right) 
 \label{eq-17}
\eea 
   in which we have introduced the  dimensionless variable $\lambda_{a} = \theta_{a}/T$      having the   effective temperature
\bea
  \theta_a \equiv \frac{ \hbar \upsilon_{\trm{F}}}{2 a k_B}.
   \label{eq-18}
\eea 

\subsection{$T \gg \theta_{a} $}

If we take the Graphene temperature to be near the room temperature, $T \approx 300 K $, then the assumption  $\theta_{a} \ll T \approx 300 $ corresponds  to a configuration of  two parallel adatom-chains with a separation distance e.g. in the order of few micrometers, see Eq. \eqref{eq-18} (with $ \upsilon_{\trm{F}} \approx 10^6, \hbar \approx 10^{-34}$ and $k_B \approx 10^{-23})$. An asymptotically suitable expression for the zero-point energy for  high temperatures, would be provided generally by applying the familiar \tit{heat kernel expansion}, see e.g. \cite{CR3}. However here we use a new useful technique to obtain  an exact expression, for the zero-point energy, being asymptotically suitable for  sufficiently high temperatures (respecting the effective temperature).  First separating $l=0$ from the $l$-sum, we rewrite Eq. \eqref{eq-17} as 
 \bea
E_0(a,T)&=& 2 L \sqrt{\pi}\frac{(k_B T)^2}{\hbar \upsilon_{\trm{F}}}  \lim_{s\rightarrow 0} \frac{\partial}{\partial s}  \left(\frac{\hbar \upsilon_{\trm{F}}}{2\pi  k_B T\delta}\right)^{2s}  \nn \\
 && \hspace{1 cm} \times \int_0^{\infty} \frac{dt}{t} \frac{t^{s-1/2}}{\Gamma(s)} \left[1+2 \sum_{l=1}^{\infty}\exp \left(-t  l^2\right) \right]  \sum_{n=0}^{\infty} \exp \left[-t \lambda_a^2 \left(n+\frac12\right)^2 \right] \nn \\
 &=&  2 L \sqrt{\pi}\frac{(k_B T)^2}{\hbar \upsilon_{\trm{F}}}  \lim_{s\rightarrow 0} \frac{\partial}{\partial s}  \left(\frac{\hbar \upsilon_{\trm{F}}}{2\pi  k_B T\delta}\right)^{2s} \Bigg(\frac{\zeta_{\trm{H}} \left(2 s-1,\frac12 \right) \Gamma \left(s-\frac{1}{2}\right)}{\Gamma (s) \lambda _a^{2 s-1}}  \nn \\
 && \hspace{1 cm} + 2\int_0^{\infty} \frac{dt}{t} \frac{t^{s-1/2}}{\Gamma(s)}  \sum_{l=1}^{\infty}\exp \left(-t  l^2\right)  \sum_{n=0}^{\infty} \exp \left[-t \lambda_a^2 \left(n+\frac12\right)^2 \right] \Bigg)
 \label{eq-19}
\eea 
where in the third line,  we have utilized the known Hurwitz zeta function
\be
\zeta_{\trm{H}}(s,r) =\sum_{n=0}^{\infty} (n+r)^{-s}.
\label{eq-20}
\ee
Then we use the following formula from Ref. \cite{math1};
\bea
&&\sum_{n=0}^{\infty} \exp\left[-c\left(n+d \right)^2 \right]=\sum_{n=0}^{\infty} \frac{(-1)^n}{n!}c^n \zeta_{\trm{H}}(-2n,d)  +\frac12 \sqrt{\frac{\pi}{c}}\nn \\ 
&& \hspace{4cm}+ \sqrt{\frac{\pi}{c}}\cos(2\pi d)\sum_{n=1}^{\infty} \exp\left[-\frac{n^2\pi^2}{c^2}\right]; \ \  c>0, \ d\geq 0.
\label{eq-21}
\eea
 For $d=1/2$, the above equation would be simplified to 
\bea
\sum_{n=0}^{\infty} \exp\left[-c\left(n+\frac12 \right)^2 \right]=   \frac12 \sqrt{\frac{\pi}{c}} - \sqrt{\frac{\pi}{c}} \sum_{n=1}^{\infty} \exp\left[-\frac{n^2\pi^2}{c^2}\right],
\label{eq-22}
\eea
where we have used $\zeta_{\trm{H}}(-2n,1/2)=0 \ ; \  n=0,1,...$, see e.g. \cite{math2}. Now, by applying Eq.\eqref{eq-22} to the $n$-sum in  the fourth line of  Eq. \eqref{eq-19}, and after some calculations we find
 \bea
&& E_0(a,T)  =  2 L \sqrt{\pi}\frac{(k_B T)^2}{\hbar \upsilon_{\trm{F}}}  \lim_{s\rightarrow 0} \frac{\partial}{\partial s}  \left(\frac{\hbar \upsilon_{\trm{F}}}{2\pi  k_B T\delta}\right)^{2s-1}   \nn \\
 && \hspace{3 cm} \times  \Bigg(\frac{\zeta_{\trm{H}} \left(2 s-1,1/2\right) \Gamma \left(s-\frac{1}{2}\right)}{\Gamma (s) \lambda _a^{2 s-1}}+\frac{\sqrt{\pi }}{\lambda_a}\frac{ \Gamma (s-1)}{\Gamma (s)}  \zeta_{\trm{R}} (2 s-2)  \nn \\
 && \hspace{4 cm} -\frac{4 \sqrt{\pi }}{\lambda _a \Gamma(s)} \sum_{l,n=1}^{\infty}  \left(\frac{l}{n \pi/\lambda_a}\right)^{-s+1} K_{-s+1}\left(\frac{2 l n\pi}{\lambda_a}\right)\Bigg)
 \label{eq-23}
\eea 
in which,  $\zeta_{\trm{R}}$ denotes the well known Riemann zeta function;
\be
\zeta_{\trm{R}}(s) =\sum_{n=1}^{\infty} n^{-s}.
\label{eq-24}
\ee 
and we have used  the integral relation
\be
\int_0^{\infty}  t^r \exp\left[-x^2 t-y^2/t \right]dt=2(x/y)^{-r-1}K_{-r-1}(2xy). 
\label{eq-25}.
\ee
 with $K$  as  a Bessel function of the second kind.  Then, after some calculations,
 the zero-point energy \eqref{eq-23} takes the form
\bea
E_0(a,T)=- 2 L \frac{( k_{\trm{B}} T)^2}{\hbar  \upsilon_{\trm{F}}} \left(\frac{2 \pi \zeta '(-2)}{\lambda _a  }+\frac{\pi   \lambda _a}{12  } +  4\sum_{l,n=1}^{\infty}  \frac{ l}{n} K_1\left(\frac{2 l n \pi }{\lambda _a}\right)\right)
\label{eq-26}
\eea
in which, ``prime'' denotes  the differentiating with respect to the argument. To obtain the Casimir energy, one  must subtract the contribution of  free (2-dimensional) space from Eq. \eqref{eq-26}. To obtain the zero-point energy contribution of the free (i.e. unbounded) space,  one just need to turn the discrete modes to continuous ones (see e.g. \cite{CR3}), which for the zero-point energy \eqref{eq-15}, we have
 \be
E^{\trm{free}}_0 (a,T) = -2 k_B T \sum_{l=-\infty}^{\infty} \int_{-\infty}^{\infty} \frac{L}{2\pi} d k_y \int_{-\infty}^{\infty} \frac{a}{2\pi} d k_x  \ln\left[\left(\frac{2\pi k_B T}{\hbar \upsilon_{\trm{F}}}\right)^2  l^2+ k_x^2+ k_y^2\right]. 
\label{eq-27}
 \ee
Using similar calculations, one  can show that the zero-point contribution \eqref{eq-27} is exactly equal to the first term of  Eq. \eqref{eq-26},
\bea
E^{\trm{free}}_0 (a,T) = -  L \frac{( k_{\trm{B}} T)^2}{\hbar  \upsilon_{\trm{F}}} \frac{4 \pi \zeta '(-2)}{\lambda _a  }
\label{eq-28}
\eea
 Thus  subtracting this term, we find the fermionic   Casimir energy of the Graphene:
\bea
E_{\trm{C}}(a,T)=- L\left(\frac{\pi  k_{\trm{B}}   T}{12 a}+\frac{8  k_{\trm{B}}^2  T^2}{ \hbar  v_F}\sum_{n,l=1}^{\infty} \frac{l}{n} K_1\left(4 l n \pi a\frac{ k_{\trm{B}} T}{\hbar  v_F}\right)\right)
\label{eq-29}
\eea
see Eq. \eqref{eq-18}. So the  fermionic Casimir force between adatom-chains, $F_{\textrm{C}}\equiv - \frac{\partial}{\partial a}  E_C $, would be obtained  as
\bea
 F_{\textrm{C}} (a,T)  =- L \Bigg(\frac{\pi  k_{\trm{B}}   T}{12 a^2}+\frac{16 \pi  k_{\trm{B}}^3   T^3}{\hbar ^2 v_F^2}  \sum_{n,l=1}^{\infty} l^2 \left[K_0\left(4 l n \pi a\frac{ k_{\trm{B}} T}{\hbar  v_F}\right)+K_2\left(4 l n \pi a\frac{ k_{\trm{B}} T}{\hbar  v_F}\right)\right]\Bigg)
\label{eq-30}
\eea
 Note that,  Eqs. \eqref{eq-29} and \eqref{eq-30} are exact expressions for the Casimir energy and the Casimir force, however,  they provide  asymptoticly explicit expressions for temperatures sufficiently larger than the effective temperature. In fact for sufficiently large temperatures (in comparison to the effective temperature \eqref{eq-18}, the Casimir energy  \eqref{eq-29} and the Casimir force \eqref{eq-30}, can be approximated with their first term.  The above useful  approach  can be generalized directly to find asymptotic expressions for the    Casimir energy  in spaces with arbitrary dimensions.  Note also  that the above equation represents an attractive Caimir interaction between two (parallel) chains of adatoms, which is in agreement  to the result of Ref. \cite{IA} for the  Casimir interaction between two  single adatoms.

\subsection{$T \ll \theta_{a} $}

At room temperature, the assumption $T \ll \theta_{a}$ corresponds to a configuration of  two parallel adatom-chains with a separation distance e.g.   in the order of few nanometers.  In order to find an asymptotically explicit expression for the zero-point energy \eqref{eq-15}  for small temperatures, we apply  Eq. \eqref{eq-21} with $d=0$,
\bea
\sum_{n=0}^{\infty} \exp\left[-c n^2 \right]= \frac12+  \frac12 \sqrt{\frac{\pi}{c}} + \sqrt{\frac{\pi}{c}} \sum_{n=1}^{\infty} \exp\left[-\frac{n^2\pi^2}{c^2}\right],
\label{eq-31}
\eea
where we have used $ \zeta_{\trm{H}}(0,0)=\frac12, \  \zeta_{\trm{H}}(-2n,0)=0 \ ; \  n=1,2,...$, see e.g. \cite{math2}. Note that the above equation is equivalent just to the familiar Poisson sum formula, see e.g. \cite{math3}. Now applying  the formula  \eqref{eq-31} to the $l$-sum in the fourth line of  Eq. \eqref{eq-19}, and using Eqs. \eqref{eq-20} and \eqref{eq-25},  one finds
  \bea
&&E_0(a,T) =  2\pi L  \frac{(k_B T)^2}{\hbar \upsilon_{\trm{F}}}  \lim_{s\rightarrow 0} \frac{\partial}{\partial s}  \left(\frac{\hbar \upsilon_{\trm{F}}}{2\pi  k_B T\delta}\right)^{2s} \Bigg(\frac{\zeta \left(2 s-2,\frac{1}{2}\right) \Gamma (s-1)}{\Gamma (s) \lambda _a^{2 s-2}}+  \nn \\
 && \hspace{2 cm} + \frac{4}{\Gamma(s)}\sum_{n=0}^{\infty}\sum_{l=1}^{\infty}\left(\frac{(2n+1 )\lambda_a}{2 l \pi}\right)^{- s+1} K_{-s+1}\left[ l \left(2n+ 1\right)\pi \lambda_a\right]\Bigg)
 \label{eq-32}
\eea 
Then after some similar calculations as before, the above equation takes the form
\bea
E_0(a,T)= 2\pi L \frac{( k_{\trm{B}} T)^2}{\hbar  \upsilon_{\trm{F}}} \Bigg(\frac{3 \pi   \lambda _a^2 \zeta '(-2)}{2 }+ 2  \lambda _a \sum_{n=0}^{\infty}\sum_{l=1}^{\infty} \frac{2n+1}{l }  K_1\left[ l \left(2 n+1 \right) \pi  \lambda _a\right]\Bigg)
\label{eq-33}
\eea
Then subtracting the contribution of free space, Eq. \eqref{eq-28}, from the above equation, we obtain the Casimir energy
\bea
&&E_{\trm{C}}(a,T)=-L\Bigg(\frac{3 \zeta_{\trm{R}}(3)}{16\pi} \frac{  \hbar   v_F}{a^2}+\frac{4 \zeta_{\trm{R}}(3)}{ \pi} \frac{ (k_{\trm{B}} T)^3 a }{(\hbar v_F)^2} \nn \\
 && \hspace{4cm} -\frac{2 k_{\trm{B}} T}{a } \sum_{n=0}^{\infty}\sum_{l=1}^{\infty} \frac{(2 n+1) }{l}K_1\left(l (2 n+1) \pi\frac{  \hbar  v_F}{2 a k_{\trm{B}} T}\right)  \Bigg)
\label{eq-34}
\eea
where we have used 
\be
\zeta'_{\trm{R}}(-2)=-\frac{\zeta_{\trm{R}}(3)}{4\pi^2}; \ \ \ \zeta_{\trm{R}}(3) \approx 1.2 
\label{eq-35}
\ee
see e.g. \cite{math2}. To find asymptotic expression for the Casimir energy for sufficiently small temperatures (in comparison to the effective temperature \eqref{eq-18}), the last term of   \eqref{eq-34} can be neglected.
 
\subsection{Influence on the  initial tension of a suspended Graphene}
 It has been experimentally demonstrated that a  Graphene sheet  suspended over a trench, possesses a rather large initial tension, which is arisen from e.g. the strong  Van der Waals interactions between the Graphene   and the sidewalls of the trench, and/or from the external forces during the fabrication of the Graphene sheet, see e.g. \cite{GV1,GV2,GV3,GV4,GT1,GT2,GT3}. Such a large built-in tension  plays an important role in the resonance frequencies of the Graphene resonators, see e.g. \cite{GV1,GV2,GV3,GV4}. Here we want to show that the the fermionic Casimir forces can have considerable influence  on the the mentioned built-in tension of a suspended Graphene. In fact  considering a suspended Graphene sheet with an initial  tension ,  and with two parallel adatom-chains as given  in  Fig. \ref{AP}, then the  fermionic Casimir force between  the adatom chains, can be considered as a (quantum) contribution term to the initial tension of Graphene. Note however that the above consideration is  meaningful, if the separation distance between adatom-chains is sufficiently smaller than sidelengths of the Graphene sheet. As a result, the change in the  initial tension (per unit length) of   Graphene sheet  due to the fermionic Casimir forces between (per unit length) of adatom chains, at room  temperature, using Eq. \eqref{eq-34}, can be written as
  \be
\Delta \tau (a) \approx \frac{3 \zeta_{\trm{R}}(3)}{8\pi} \frac{  \hbar   v_F}{a^3}; \ \  a < \trm{ few nanometer}
 \label{eq-36}
 \ee
in which $\tau$ is the initial tension of the suspended Graphene sheet. The above equation obviously means that the attractive Casimir force between the adatom chains, increases  the initial tension of the Graphene, as physically expected. For $a= 1 \trm{nm}$ in Eq. \eqref{eq-36} one finds a value of the order of $10^{-2} N/m$ for $\Delta \tau$, which is considerably noticeable in comparison to the experimental values of the order of $10^{-1}$- $10^{-4} N/m$,  given in Refs. \cite{GT1,GT2,GT3}, for the initial tension  of the Graphene sheet with sidelengths in the order of few micrometers.

\end{document}